# A Proof-of-Concept Device-to-Device Communication Testbed

Vibhutesh Kumar Singh
Wirocomm Research Group
Department of ECE
IIIT-Delhi, India
vibhutesh.k.singh@ieee.org

Hardik Chawla
Department of EEE&I
BITS Pilani Goa Campus, India
hchawla94@gmail.com

Vivek Ashok Bohara
Wirocomm Research Group
Department of ECE
IIIT-Delhi, India
vivek.b@iiitd.ac.in

*Abstract*— **This paper presents the design and development of proof-of-concept Device-to-Device (D2D) Communication testbed. This testbed also seeks to address the design issues involved in the implementation of a D2D network in a realistic scenario. The performance of this testbed has been validated by emulating a Cellular network consisting of a Base Staion (BTS) and many D2D devices in its proximity. The devices and the BTS coordinate and communicate with each other to select the optimum communication range, mode of communication and transmit parameters. Through the experimental results it has been shown that the proposed testbed has a communication radius of 120m and a D2D communication range of 62m with over 90% efficiency.**

## I. Introduction

Every natural calamity reminds us of our heavy dependence on infrastructure for information dissemination, and the need of easily deployable emergency communication services. In any catastrophic natural calamity, communication services are plagued by last mile connectivity issues and destruction of infrastructure. For instance, after Nepal earthquake [1] and Siachun earthquake in China [2], most of cellular infrastructure became inoperable within minutes.

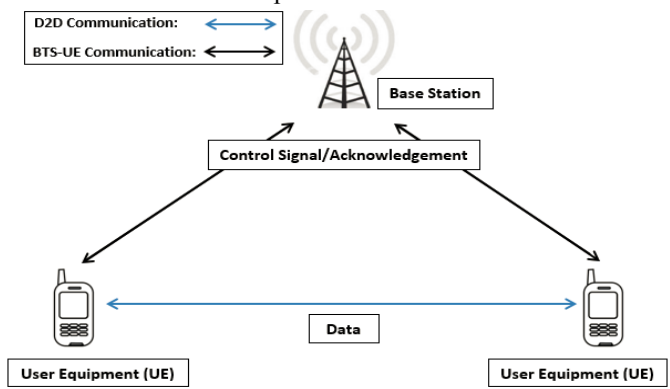

*Fig. 1. A Typical D2D Communication Scenario, With a Relay*

This motivates the use of a cellular service that is minimally dependent on network infrastructure. D2D communication is one such service. In a typical D2D communication scenario, the user equipment (UE) within the D2D range exchange data directly through the D2D link circumventing the BTS, thus achieving high data rate, and low latency as compared to conventional cellular services. It also reduces the load on the network and provides robustness against infrastructure failures [3]. Additionally, inclusion of relays in D2D network enhances the capacity and coverage of networks [4].

Fig. 1. And Fig. 2. represents typical D2D scenarios with and

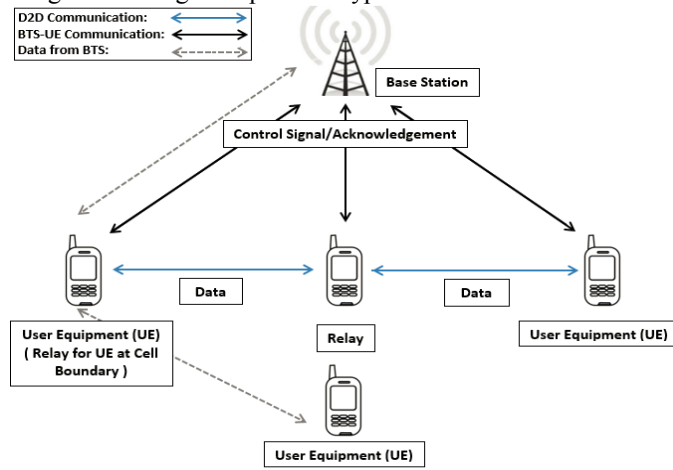

*Fig. 2. A Typical D2D Communication Scenario, With a Relay*

without, a relay. Normally, if the devices are in proximity, relaying is not required and, BTS allocates resources for direct D2D communication. However, if the devices cannot form a D2D pair (due to distance, and/or unfavorable channel conditions) the information is relayed through another device which is in proximity to both the devices. This enhances the range of D2D communication and maintains efficiency.

In this work we have presented a simple, low cost, readily deployable D2D communication testbed. It emulates a portable BTS and duplex D2D transceivers which can act as relays on request. The results obtained can thus be extended to develop future 5G networks of which D2D communication is an integral part [5]. This proof-of-concept wireless network has been implemented using IEEE802.15.4 standard device [6] and 2.4 GHz RF Transceivers (CC2500) [7] which are battery powered, and known for their sparse energy consumption.

## II. Experimental Setup

This section discusses the experimental setup of our testbed. Atmega328p [8] based microcontroller board was used for modeling user equipments (UEs) or devices. Separate transceivers were used for D2D and BTS-UE communication. CC2500 based transceiver [7] was used for D2D communication due to its high power efficiency and small communication range. Zigbee transceivers [6] were used for

BTS-UE communication because of communication range and device addressing capability. The BTS was modeled using serial communication enabled software [9]. The presented testbed operates in the ISM bands of 2.4GHz. The Forward Error Correction (FEC) Coding feature was enabled to reduce the packet errors. With FEC the obtained data rate is 57.6 k Baud.

The BTS on the basis of Received Signal Strength Indicator (RSSI) value of packets decides the following:
(a) Whether D2D communication is feasible.
(b) If (a) is true, then whether it should take place through a direct link or through a relay.

## III. EXPERIMENTAL RESULTS AND ANALYSIS

This section describes the experimental results and analyses for D2D testbed. To test BTS-UE link, we have transmitted 50 packets from different distances and have accounted the RSSI value of each packet, which gives us an estimate of the communication range and efficiency of the link:

A. *RSSI vs Distance:* As the distance between the BTS and UE increases, the RSSI value decreases and the variance of RSSI increases. It is primarily due to the path loss in the wireless channel and decreasing SNR at larger distances. The RSSI value is used to estimate the approximate distance between the BTS and UE within the cell. Based on this model, the BTS can predict the possibility of D2D communication between different devices by setting a threshold RSSI value. RSSI based localization [10] is an established method to localize various objects within a locality and could be accurately used to estimate the radial distances.

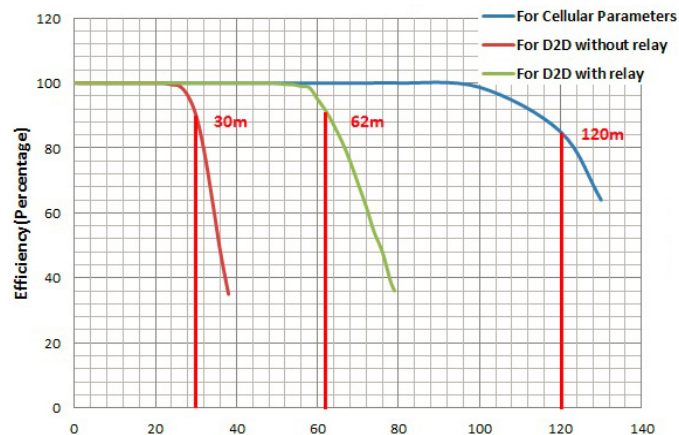

*Fig. 3. Variation of Packet Efficiency with RSSI of base station*

B. *Efficiency vs. Distance*: Similar to experimental results obtained in [11] we observed that as RSSI decreases the packet drop count increases. Results from Fig. 3. helped us estimate the cell radius, and the range of D2D communication with/ without relay. From the data collected, we conclude the cell radius is approximately 120m which caters to an area of 0.045km$^2$ with an efficiency threshold of 85%. We also observed that the D2D range without relay which was 30m with efficiency threshold of 90% got extended by almost twice to 62m when relays were incorporated. Additionally, a steeper drop in packet efficiency in case of D2D without relay was observed as compared to the case of D2D with relay. Here, efficiency is: *(Number of packets received correctly/Number of packets transmitted) \*100*.

C. *Power Consumption metric*

|  | **Power and Time** |
|---|---|
| Power Consumption (D2D Feature ON) | **385.2mW** |
| Power Consumption (D2D Feature OFF) | **234mW** |
| Active time (D2D Feature ON) | **13.75hr** |
| Standby-time (D2D Feature Off) | **22.64hr** |

*Table 1. Power consumption and Operation time metric of testbed in various operation modes*

Table 1. shows the power consumption of modules with in both modes i.e., with D2D feature ON and OFF. Standby times were measure using a 5.3Wh 3.6V battery as reference.

## IV. CONCLUSIONS

An experimental testbed for D2D communication was designed and implemented using IEEE802.15.4 and 2.4 GHz

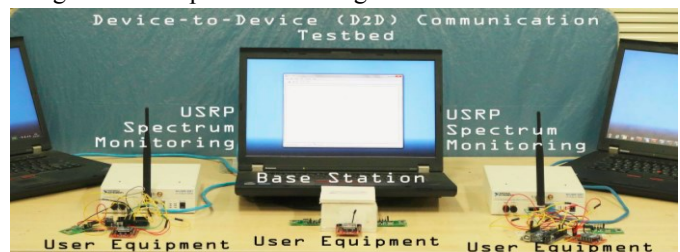

*Fig. 4. Device-to-Device (D2D) Communication testbed*

RF Transceivers. The results and analysis indicate that the testbed is highly scalable and can be implemented in a real world scenario by following a similar approach. Highly portable architecture of this testbed enables it to be readily deployed in a disaster stricken area to provide immediate communication services. The range of the testbed can be further extended by incorporating more devices, large power amplifiers and powerful antennas. By using sophisticated algorithms and Software Defined Radio platforms the concepts of network coding and cognitive radio networks can be applied to make the system more advanced and secure.

## REFERENCES

[1]. K. Goda, T. Kiyota, R. Pokhrel, G. Chiaro, T. Katagiri, K. Sharma and S. Wilkinson, 'The 2015 Gorkha Nepal Earthquake: Insights from Earthquake Damage Survey', Frontiers in Built Environment, vol. 1, 2015.
[2]. Gupta, Harsh, et al. "The Disastrous M 7.9 Sichuan Earthquake of 12 May 2008." *Journal of the Geological Society of India* 72.3 (2008): 325-330.
[3]. Y. Xu, R. Yin, T. Han and G. Yu, 'Dynamic resource allocation for Device-to-Device communication underlaying cellular networks', *Int. J. Commun. Syst.*, vol. 27, no. 10, pp. 2408-2425, 2012.
[4]. Amate, Ahmed Mohd. *D2D Communication and Multihop Transmission for Future Cellular Networks*. Diss. University of Hertfordshire, 2014.
[5]. Boccardi, Federico, et al. "Five disruptive technology directions for 5G." Communications Magazine, IEEE 52.2 (2014): 74-80.
[6]. Pro, XBee. "Data sheet Digi International Inc."
[7]. Texas Instruments. "Low-power 2.4 GHz RF transceiver." *CC2500 datasheet* (2010).
[8]. Atmel Coorporation, "Atmel ATmega328P Datasheet".
[9]. X.C.T.U. "Test Utility Software: User's Guide."*Digi International 2015*.
[10]. Sugano, Masashi, Tomonori Kawazoe, Yoshikazu Ohta, and Masayuki Murata. "Indoor localization system using RSSI measurement of wireless sensor network based on ZigBee standard." *Target* 538 (2006): 050.
[11]. Holland, Matthew M., Ryan G. Aures, and Wendi B. Heinzelman. "Experimental investigation of radio performance in wireless sensor networks." In *Wireless Mesh Networks, 2006. WiMesh 2006. 2nd IEEE Workshop on*, pp. 140-150. IEEE, 2006.